\documentclass[%
reprint,
bibnotes,
amsmath,
amssymb,
aps,
floatfix,
]{revtex4-2}

\usepackage{graphicx}
\usepackage{subfigure}
\usepackage{bm}
\usepackage{amsmath}
\usepackage[colorlinks=true,citecolor=blue,urlcolor = blue,linkcolor= blue]{hyperref}
\usepackage{epstopdf}
\usepackage{color,soul}
\usepackage[toc,page]{appendix}
\usepackage{soul}
\usepackage{color}

\makeatletter
\AtBeginDocument{\@ifpackageloaded{natis}{\ifNAT@numbers\if@filesw\immediate\write\@auxout{\string\global\string\NAT@numberstrue}\fi\fi}{}}
\makeatother

\begin{document}

\renewcommand{\thefigure}{\ \arabic{figure}}

\title{Correlation between vibrational anomalies and emergent anharmonicity of local potential energy landscape in metallic glasses}

\author{Zeng-Yu Yang}
\affiliation{State Key Laboratory of Nonlinear Mechanics, Institute of Mechanics, Chinese Academy of Sciences, Beijing 100190, China}
\affiliation{School of Engineering Science, University of Chinese Academy of Sciences, Beijing 100049, China}

\author{Yun-Jiang Wang}
\email{yjwang@imech.ac.cn}
\affiliation{State Key Laboratory of Nonlinear Mechanics, Institute of Mechanics, Chinese Academy of Sciences, Beijing 100190, China}
\affiliation{School of Engineering Science, University of Chinese Academy of Sciences, Beijing 100049, China}

\author{Alessio Zaccone}
\email{alessio.zaccone@unimi.it}
\affiliation{Department of Physics ``A. Pontremoli'', University of Milan, via Celoria 16, Milan 20133, Italy}
\affiliation{Department of Chemical Engineering and Biotechnology, University of Cambridge, Cambridge CB3 0AS, UK}
\affiliation{Cavendish Laboratory, University of Cambridge, Cambridge CB3 0HE, UK}

\date{\today}

\begin{abstract}

The boson peak (BP) is a universal feature in the Raman and inelastic scattering spectra of both disordered and crystalline materials. Here, through a set of atomistically-resolved characterizations of metallic glasses, we uncover a robust inverse proportionality between the intensity of boson peak and the activation energy of excitations in the potential energy landscape (PEL). Larger boson peak is linked with shallower basins and lower activation barriers and, consequently, with emergent anharmonic sectors of the PEL. Numerical evidence from atomistic simulations indicates that THz atomic vibrations contributing the most to the BP in atomic glasses are strongly correlated with such emergent anharmonicity of PEL, as evidenced through very large values of the atomic- and mode-resolved Gr\"{u}neisen parameter found for the atomic vibrations that constitute the BP. These results provide a direct bridge between the vibrational spectrum and the topology of the PEL in amorphous solids.

\end{abstract}

\maketitle


\section{Introduction}
The spectra of atomic vibrations in disordered materials, such as glasses, have been the object of tremendous experimental, computational and theoretical efforts over the last decades. Understanding the vibrational Raman and inelastic neutron/X-ray scattering spectra of glasses is a central step for the quantitative prediction and understanding of the thermodynamic and thermal transport properties of disordered materials. While the electronic properties of disordered systems are now fundamentally understood thanks to the pioneering work of Mott~\cite{Mott}, Anderson~\cite{Anderson}, Efros, Shklovskii~\cite{Efros} and others~\cite{Meyer1996}, such a fundamental understanding for the vibrational and thermal properties of structurally disordered materials is currently missing.

The vibrational density of states (VDOS), in this context, plays a central role since it is the key factor entering the integrals in terms of which the specific heat and thermal conductivity of glasses are expressed~\cite{Zeller1971}.
Since the early 1960s, at least, experimental evidence from Raman and Brillouin scattering of glasses showed the presence of a large peak in the Debye-normalized Raman intensity (i.e. divided by Debye's law $\sim \omega^{2}$), in the THz regime~\cite{Leadbetter}. The Raman intensity of glasses at low energy is given by \cite{Shuker,Leadbetter} $I(\omega)\sim g(\omega)[n(\omega,T)+1]$, with $n(\omega,T)+1=[1-\exp{(-\hbar \omega/k_{B}T})]^{-1}$ the Bose function. Here, $g\left( \omega  \right)$ denotes the VDOS and $k_{B}$ is the Boltzmann constant~\cite{Shuker}. Note that in Raman scattering experiments, one does not measure $g(\omega)$ directly but $g(\omega)C(\omega)$, where the function $C(\omega)$ describes the coupling of the radiation with the sample. This gives an additional dependence on $\omega$ and hence the boson peak is not at the same frequency as the one found in neutron-scattering \cite{Taraskin1997}. Since the peak intensity appeared to depend on temperature $T$ according to the Bose distribution (likely because the exponential character of the latter obscures all other dependencies), this prompted researchers to believe that the boson peak is insensitive to temperature and therefore its origin must be purely ``harmonic''.

In spite of this, the early theoretical approaches to explain the boson peak vibrational glassy anomalies were based on double-well anharmonic models~\cite{Klinger,Buchenau1991,Buchenau1992,Gurevich1993,Gurevich2003}, following in the wake of Ilya M. Lifshitz's pioneering work on atomic vibrations around defects in solids \cite{Lifshitz1966}. Further experimental evidence were collected later on, revealing the profound effects of anharmonicity on the attenuation of sound waves in glasses in the GHz and THz regions, supporting the anharmonic origin of Brillouin (or Akhiezer) diffusive linewidths $\Gamma \sim q^{2}$ up to the Ioffe-Regel crossover between ballistic and diffusive propagation of vibrations~\cite{Vacher2005,Vacher2008,Ruffle2008,Baldi2014}. Similar evidence for the anharmonic damping of transverse acoustic phonons has been recently found also for metallic glasses~\cite{Wang_CUHK}.
Subsequently, the Ioffe-Regel crossover between ballistic propagation and diffusive-like $\Gamma \sim q^{2}$ (transverse) excitations has been suggested based on numerical simulations, as the possible fundamental process behind the boson peak~\cite{Tanaka}.
It has been confirmed more recently in theoretical calculations~\cite{Baggioli2019,Baggioli2020} that the crossover from a regime at low $\omega$ dominated by the real (acoustic) part of the excitation into a regime dominated by the imaginary (diffusive) part provides a universal mechanism for the boson peak. The origin of the diffusive linewidth being traced back to anharmonicity~\cite{Akhiezer}, this mechanism is able to provide an explanation to the many recent observations of boson peak in the spectra of perfectly ordered or minimally-disordered crystals~\cite{Jezowski,Tamarit2017,Tamarit2019,Douglas2019}.

In spite of these many conceptual and experimental evidences suggesting an important role of anharmonicity as the driving factor for the phonon decoherence leading to Ioffe-Regel crossover and the boson peak, there is no doubt that the current dominant paradigm to explain the boson peak is based on the idea of ``dissipationless''  or ``harmonic'' disorder as manifested e.g. in spatially fluctuating elastic constants. In this framework, as developed in several papers by W. Schirmacher, G. Ruocco and co-workers~\cite{Schirmacher2007,Schirmacher2013} and known as ``heterogeneous elasticity theory'' (HET), the loss of phonon coherence at the Ioffe-Regel crossover has nothing to do with anharmonicity and stems uniquely from disorder.

In this paper we focus on this debate (i.e. whether the boson peak in glasses stems from ``harmonic'' or ``anharmonic'' processes), and provide an intimate connection between vibrational anomaly and emergent anharmonicity of the potential energy landscape (PEL) based on atomistic simulations.
We are able to directly quantify the harmonic/anharmonic character of each atomistically-resolved vibrational eigenmode that contributes to the boson peak, in a paradigmatic atomic glass. It is demonstrated that the eigenmodes that make up the boson peak in the THz regime are strongly related to anharmonicity, as reflected in their erratic trajectories through shallow regions of PEL.
Furthermore, huge values of the atomistically- and mode-resolved Gr\"{u}neisen parameter are found for the atomic vibrations that contribute to the boson peak. This rich evidence figures out the critical role of "anharmonic effect'' in glassy thermal anomalies.

\section{Methods}

\subsection{Molecular Dynamics}

Extensive molecular dynamics simulations are conducted via the open source code LAMMPS \cite{Plimpton1995}. Prototypical binary Cu$_x$Zr$_{100-x}$ ($x = 30, 40, 50, 60, 70$) MG models, each containing 19652 atoms, are constructed based on the many-body Finnis-Sinclair-type embedded-atom potential \cite{Mendelev2009} implemented to describe the interatomic interactions. For the model preparation process, a $NPT$ ensemble (constant number of atoms, constant pressure, and constant temperature) is utilized, and the pressure remains zero by Parrinello--Rahman barostat \cite{Parrinello1981}. The temperature is controlled through the Nos\'e--Hoover method \cite{Nose1984}. Each system with randomly distributed lattice atoms are first heated and equilibrated at 2000 K for 2 ns to achieve a fully melting state. The liquid is then quenched to the glassy state at 0 K, with cooling rates spanning multiple orders of magnitudes, i.e., from $10^9$ K/s to $10^{14}$ K/s. To prepare the inherent structures at different temperature, we further thermally relax the systems at the desired temperature with $NVT$ (constant number of atoms, constant volume and constant temperature), and then perform energy minimization using the conjugate gradient algorithm. Periodic boundary conditions (PBCs) are imposed to all the three directions. The MD time is set to be 0.002 ps.

\subsection{Single-particle intensity of boson peak}

The vibrational analysis of the glass state is performed by direct diagonalization of the Hessian matrix of the inherent structures, which correspond to local energy minima positions in the PEL. The single-particle vibrational density of states (VDOS) for the $i$th atom is defined as
\begin{equation}\label{Eq:1}
{g_i}\left( \omega  \right) = \frac{1}{{3N}}\sum\limits_j {\delta \left( {\omega  - {\omega _j}} \right){{\left| {\textbf{e}_j^i} \right|}^2}},
\end{equation}
where $N$ is the total number of atoms. $\omega_j$ represents the phonon frequency. $\textbf{e}^i_j$ denotes the polarization vector of $i$th atom in the vibrational mode characterized by $\omega_j$. The sum of ${g_i}\left( \omega  \right)$ equals to the total VDOS $g \left( \omega \right)$ of the system. Further, the reduced VDOS, i.e., the value divided by $\omega^2$ can be used to characterize the boson peak. The intensity of single-particle boson peak is thus formulated as the maximum value of the reduced VDOS, i.e., $I_\textrm{BP}^i = \max \left[ {{{{g_i}\left( \omega  \right)} \mathord{\left/ {\vphantom {{{g_i}\left( \omega  \right)} {{\omega ^2}}}} \right. \kern-\nulldelimiterspace} {{\omega ^2}}}} \right]$.

\subsection{Single-particle activation energy}

To understand the vibrational anomaly in metallic glasses in depth, it is necessary to explore the underlying topological feature of the PEL, including the energy minima and the surrounding saddle points, which have been demonstrated to be closely related to boson peak behavior ~\cite{Grigera2003}. To address this issue, the activation-relaxation technique nouveau (ARTn) \cite{Barkema1996,Malek2000} is utilized to extract the single-particle activation energy. In the framework of ARTn, an initial perturbation is introduced to a central atom and its neighbors by imposing a random small displacement vector. The magnitude of the perturbation displacement vector is fixed as 0.1 $\textrm \AA$, while the activation direction is chosen randomly.  To restrict the perturbation to a specific atom, the cutoff distance of the atom cluster is set to be 2 $\textrm \AA$, which is shorter than the first maximum of the radial distribution function of the CuZr glasses. The state is pulled towards high energy along the weakest Hessian direction. When the lowest eigenvalue of Hessian matrix is less than $-0.30$ eV/$\textrm \AA ^2$, the system is pushed towards the saddle point automatically using a Lanczos algorithm. The system is considered to converge to the saddle point state when the force on any atom is below 0.05 eV/$\textrm \AA$. Thereafter, the energy difference between the saddle point and the initial state is calculated as the corresponding activation energy of an atom's hopping. For a statistical purpose, each atom is activated for 20 times with random initial perturbation. The average activation energy of the 20 events for each atom is further used as a single-particle activation energy. It should be noted that only a few dozens of searches via ARTn do not guarantee to find the lowest saddle point on PEL from an inherent structure or a starting point. To find the low-lying saddle points, it is possibly more appropriate to consider other algorithms, e.g., the simple and robust algorithm proposed by Bonfanti and Kob to find lowest saddle point in complex energy landscapes \cite{Bonfanti2017}.

\section{Results and discussion}

\begin{figure}
  \centering
  \includegraphics[width=0.49\textwidth]{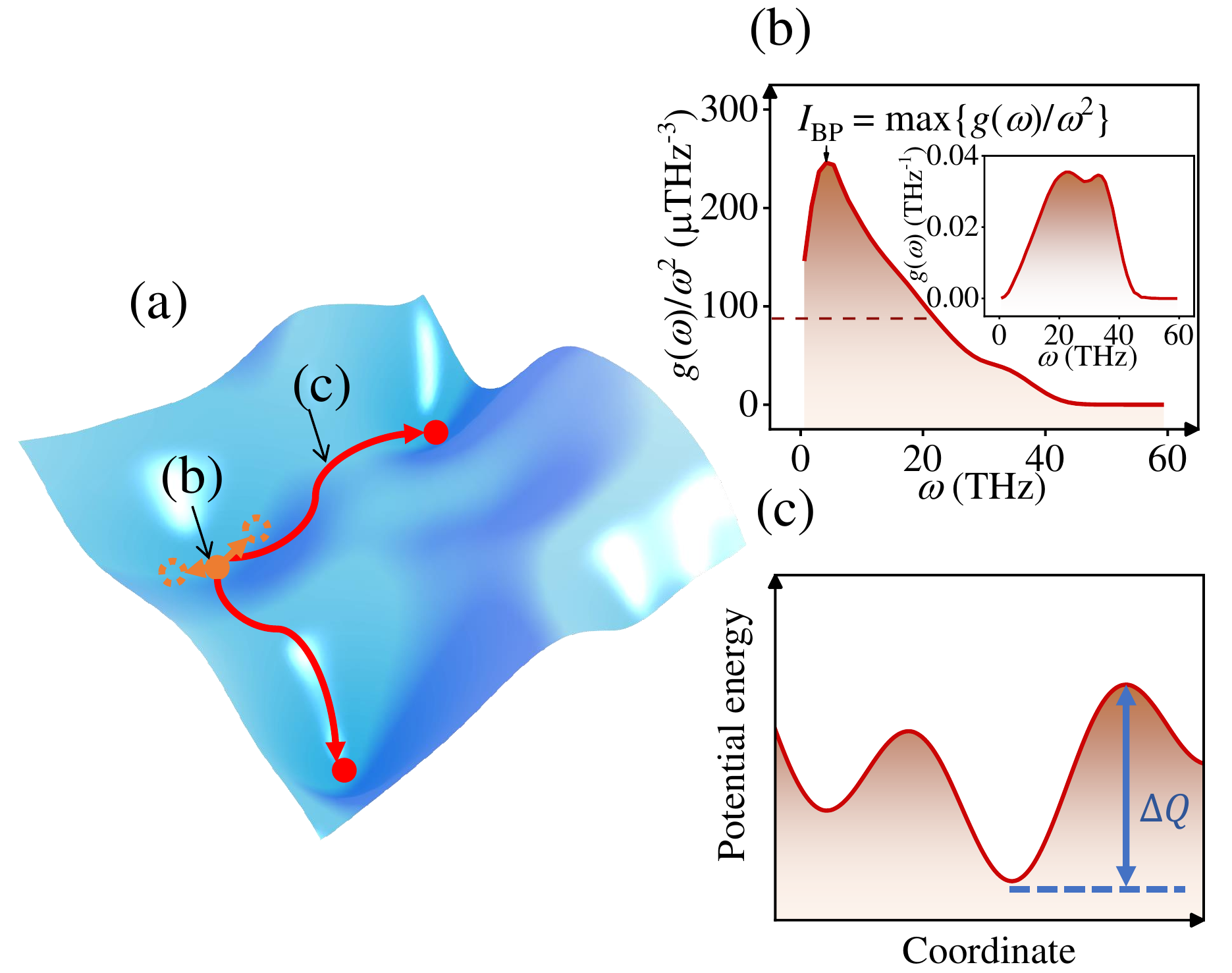}
  \caption{
  (a) Schematic of a fragment of PEL illustrating thermodynamic vibration and thermal activation between neighbouring basins. Here labels (b) and (c) refer to the other two sub-panels -- Fig.\ref{Fig:1}b and Fig.\ref{Fig:1}c, respectively. And (b) denotes the trajectory of vibration near the sub-basin while (c) represents the activation path between two adjacent sub-basins.
  (b) $\omega^2$-reduced VDOS, $g(\omega)/\omega ^2$, confirms the existence of boson peak in Cu$_{50}$Zr$_{50}$ glasses. The dashed line represents the Debye level which is $\sim$ 87.23 $\mu$THz$^{-3}$ for the present Cu$_{50}$Zr$_{50}$ glass~\cite{Yang2019}.} The inset shows the original VDOS.
  (c) Schematic of 1D activation pathway defines activation energy for a structural excitation.
  
  \label{Fig:1}
\end{figure}


\begin{figure}
  \centering
  \includegraphics[width=0.38\textwidth]{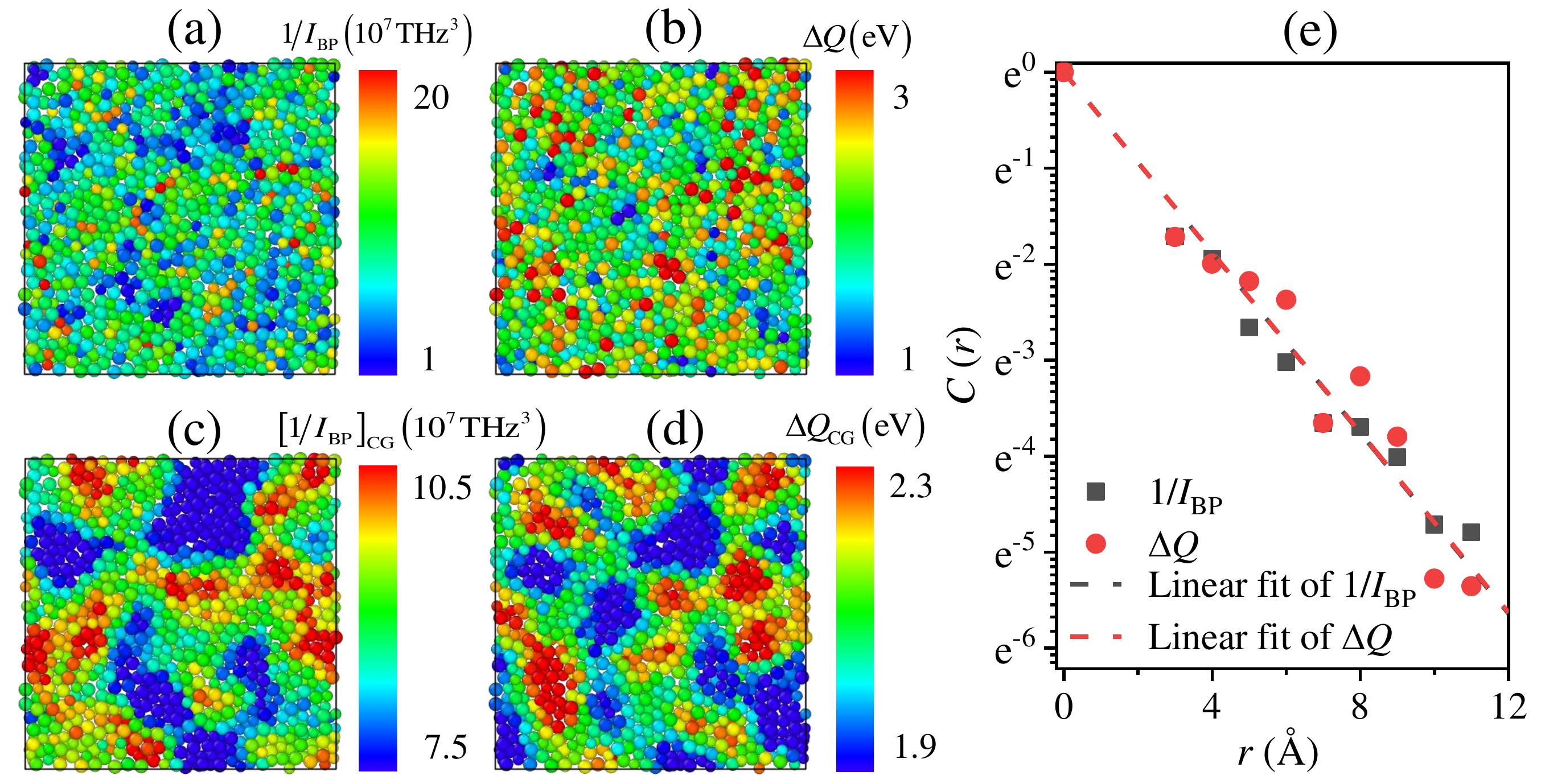}
  \caption{(a) Distribution of the activation energies for excitation of Cu atoms in two groups, i.e., the highest and lowest 10\% of the intensity of boson peak, respectively. (b) The case for Zr atoms. The activation energies shown are mean values after 20 ARTn searches.}\label{Fig:2}
\end{figure}


In order to understand what kind of excitations contribute to the boson peak, we start from the analysis of the PEL of the model binary metallic glass Cu$_{50}$Zr$_{50}$, and we study its relation to the boson peak. Fig.\ref{Fig:1}(a) shows a 3D illustration of the PEL, where the minima (or basins of the landscape) represent the inherent structures, while saddle points (or hills of the landscape) characterize the dynamical bottlenecks on the pathways of structural excitations and relaxations \cite{Fan2014,Fan2015,Fan2017}. As sketched in Fig.\ref{Fig:1}(a), the short-time vibration around a minimum or valley is shown by the orange trajectory. The long-time transition from one local energy minimum to a neighbouring one is characterized by the red line.
Clearly, the short-time vibrations within a single basin are mainly ``harmonic'', whereas the red trajectories are associated with strongly anharmonic eigenmodes. The goal of the subsequent analysis is to disentangle the relative prevalence of these two types of excitations among those that form the boson peak in the VDOS.

In amorphous materials, the existence of extra low-energy modes in excess with respect to the Debye law $\sim \omega^{2}$, defines the boson peak anomaly, which can be quantified by the maximum value of the $\omega^2$-normalized VDOS. Fig.\ref{Fig:1}(b) gives the phonon features for the Cu$_{50}$Zr$_{50}$ metallic glass which confirms the existence of the boson peak anomaly at low frequencies. It shows the position of the boson peak at a frequency of nearly 5 THz. This is in accord with experiments \cite{Li2008,Wangwh2019} which show that the energy of boson peak in Zr based metallic glasses is usually $\sim$ 5 meV.
To our knowledge, how to understand this phenomenon from the perspective of its relation to the PEL is an interesting and open question.

Figure\ref{Fig:1}(c) shows the energy difference between the saddle point and the initial local minimum, which is defined as the activation energy and is used to quantify structural excitations \cite{Ding2016,Fan2020,Wangqi2020} .
On the basis of atomistic molecular dynamics (MD) simulations, we have access to both sets of information, i.e. the activation energies in the PEL and the eigenmodes, \emph{in an atomistically resolved way}. In other words, for each atom we can extract its contribution to the vibration spectrum and to the BP, as well as its ramblings through the PEL and the activation energies that it goes through.

Hence, to seek quantitative correlations between eigenmodes and activation energies, the particle-level intensity of boson peak $I_\textrm{BP}$ and the activation energy $\Delta Q$ are calculated for each atom. The iso-line plot of single-particle activation energies as a function of the inverse intensity of boson peak is given in Fig. S1 in the Supplemental Materials (SM) \cite{sm}. Even though the correlation is somewhat broad in terms of the raw data especially for the case of Zr atoms, a qualitative trend is clear: $\Delta Q$ increases with decreasing $I_\textrm{BP}$. In the pioneering work by Manning and Liu \cite{Manning2011}, soft spots are identified via participation fraction of atoms in low-frequency vibrational modes. Here, our calculation is in accord with Ref. \cite{Manning2011} and other works which demonstrated that atoms participating preferentially in soft modes are prone to undergo shear transformations under thermal and/or mechanical stimuli \cite{Richard2021,Ding2014,Yang2021,Fan2021}. In Fig.\ref{Fig:2}, we compare $\Delta Q$  for the group of atoms with the 10\% lowest and the 10\% highest $I_\textrm{BP}$ values in Cu$_{50}$Zr$_{50}$. The atoms which more prominently contribute to the BP effectively experience a lower magnitude of activation energy in their dynamics, and, thus, are more susceptible to structural rearrangement under thermal and/or mechanical stimuli. This is consistent with the early experimental work which demonstrated that loose atoms mainly contribute independent localized vibrational modes with boson peak frequency~\cite{Tang2005}. Also, importantly, their motions are correlated with anharmonicity, as shallow activation basins are obviously linked to larger anharmonicity, whereas deep valleys and steep barriers are related to harmonic-type dynamics~\cite{Kittel,Krausser2017}.


\begin{figure}
  \centering
  \includegraphics[width=0.42\textwidth]{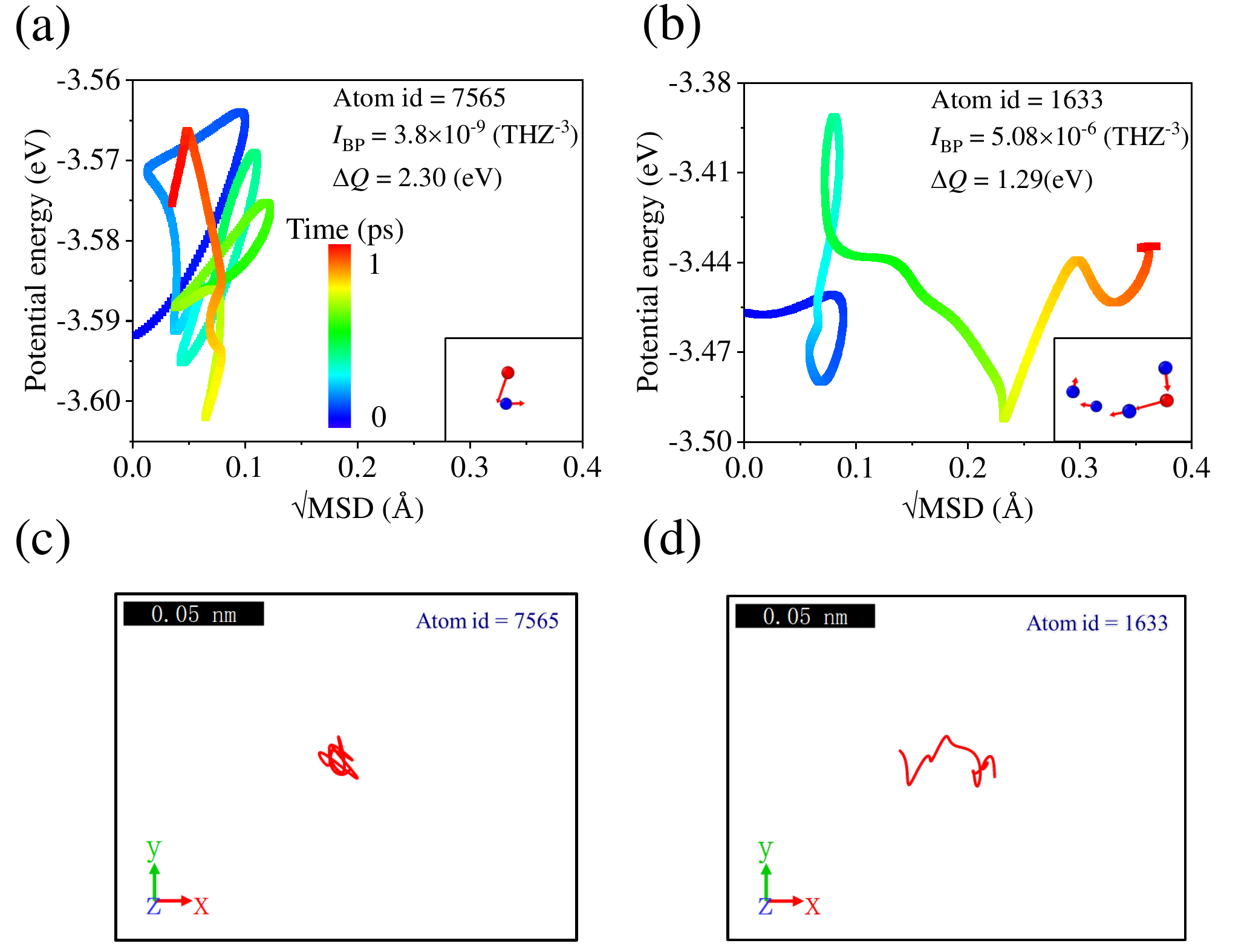}
  \caption{Pearson's correlation coefficient between boson peak intensity and activation energy as a function of spatial coarse-graining length. Best correlation achieves with $L = 5.9$ $\textrm \AA$.}\label{Fig:3}
\end{figure}

\begin{figure}
  \centering
  \includegraphics[width=0.38\textwidth]{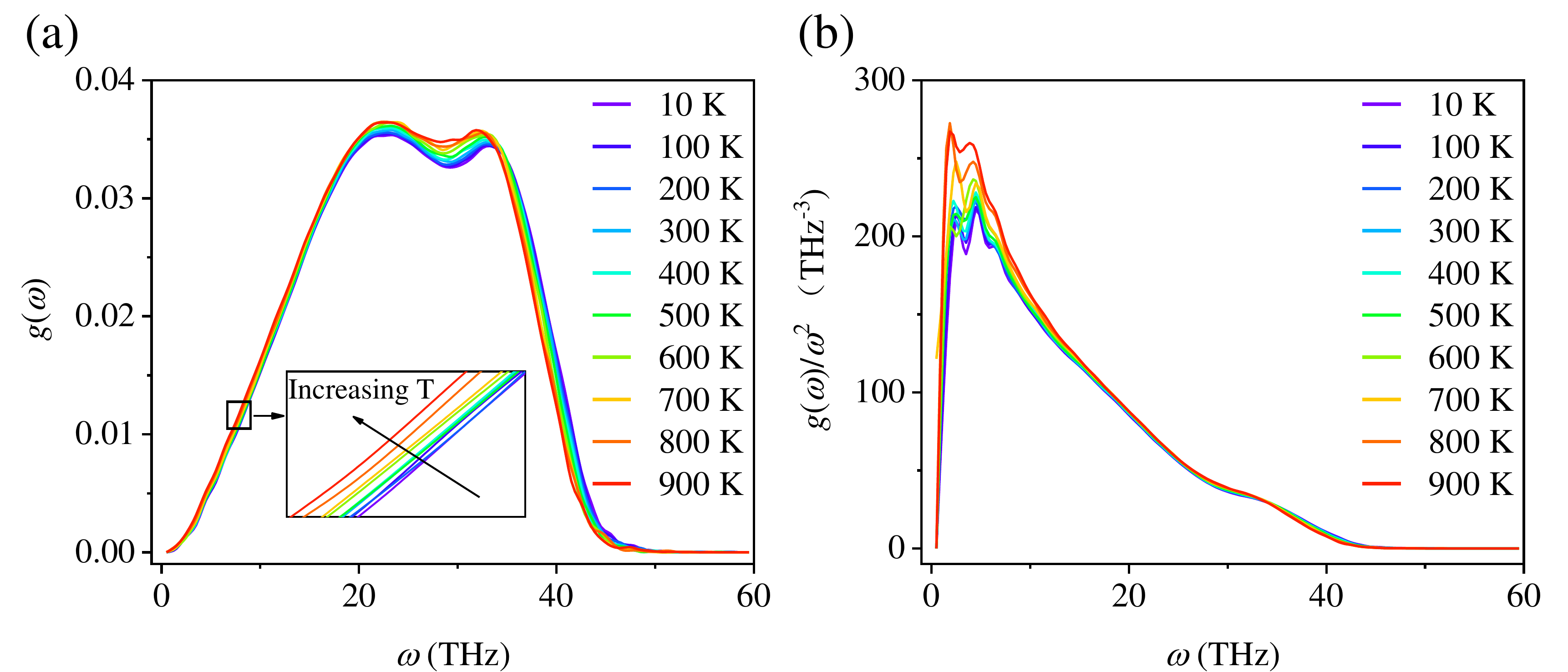}
  \caption{Boson peak correlates inversely with the difficulty of the thermally activated structural excitation in glass.
  (a) Correlation between the spatially coarse-grained boson peak intensity $[1/I_\textrm{BP}]_\textrm{CG}$ and the activation energy $\Delta Q_\textrm{CG}$ with coarse-graining length $L$ = 5.9 $\textrm \AA$. The color indicates the number density of atoms.
  (b) Numerical coarse graining of the data in (a) with binning size of a hundred atoms, which reproduces a $\Delta Q \sim I_\textrm{BP}^{-1}$ law.}\label{Fig:4}
\end{figure}

Our explanation for the weak correlation shown in Fig. S1 is that the collective vibrational anomaly and the structural excitation are not exactly controlled by the first-shell local structure, but rather their structural fingerprint is embedded in several shells of the radial distribution function (RDF) via spatial correlations. We therefore plot the $I_\textrm{BP}$ vs $\Delta Q$ correlation as a function of the coarse-graining size in Fig.\ref{Fig:3}. The best correlation, quantified by the Pearson's coefficient, can be achieved with the coarse-graining length $L = 5.9$ $\textrm \AA$ for both $I_\textrm{BP}$ and $\Delta Q$. This is exactly the same length scale which yields the best correlation, thus indicating a strong link between the activation energy and the intensity of boson peak. This characteristic length corresponds to the second valley of radial distribution function (RDF) and thus contains the short as well as medium range order of Cu$_{50}$Zr$_{50}$ glass. In Fig.\ref{Fig:4}(a), we plot the activation energy versus the reciprocal of the BP intensity for the spatial coarse-graining size that yields the strongest correlation. Finally, Fig.\ref{Fig:4}(b) shows the corresponding result after numerical coarse-graining with bin size of $\sim 100$ atoms. It shows a strong, well-defined $\Delta Q \sim I_\textrm{BP}^{-1}$ scaling law. This intimate correlation between boson peak and activation energy in the PEL implies a scenario beyond short-range order for the connection between thermodynamics and dynamics in glasses \cite{Cubuk2017,Hu2018,Tong2018,Tong2020}.

\begin{figure*}
  \centering
  \includegraphics[width=0.75\textwidth]{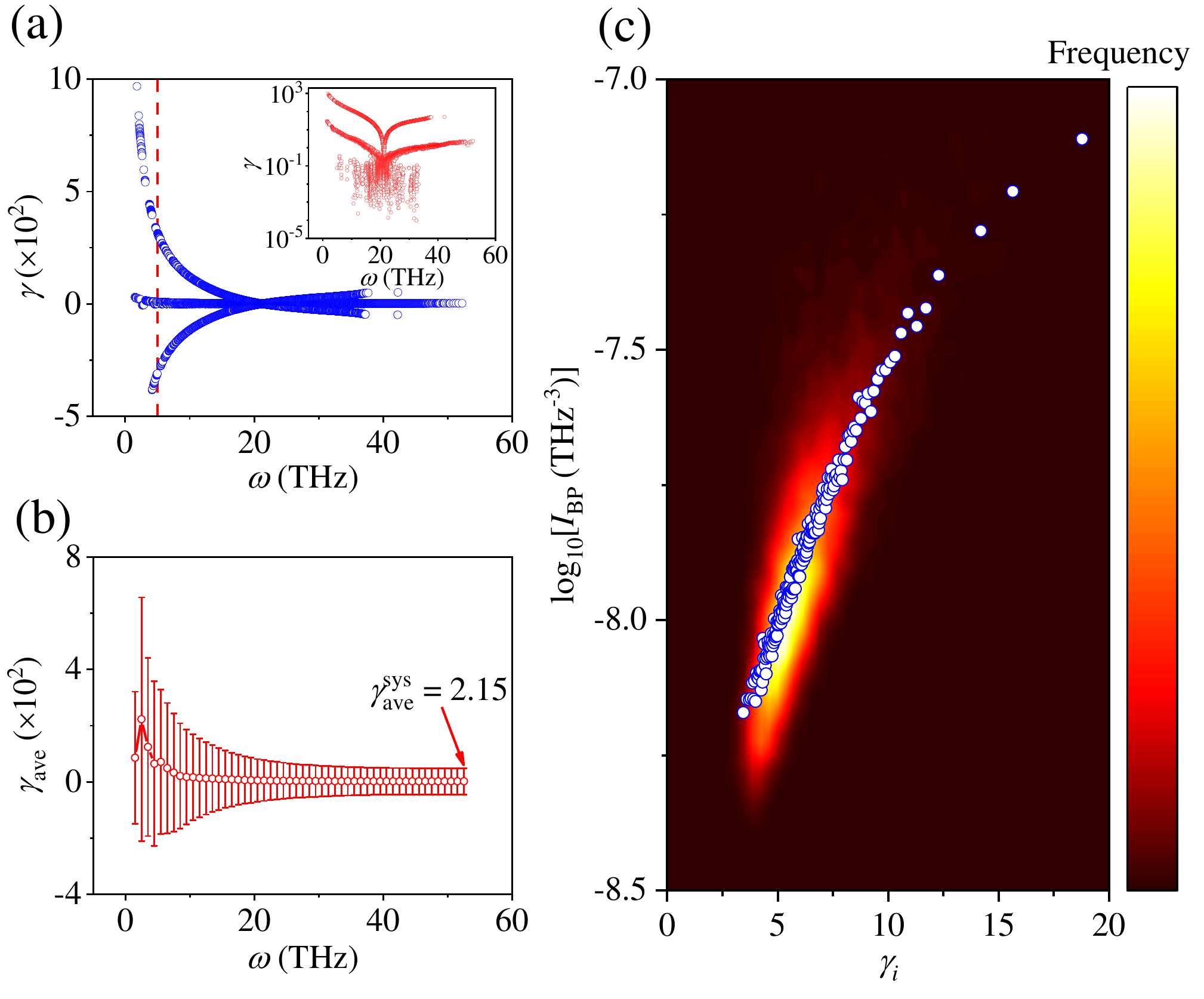}
  \caption{Spatial correlation between the single-particle boson peak and the activation energy.
  (a)--(d) Spatial distribution of $1/I_\textrm{BP}$, activation energy $\Delta Q$, as well as their coarse-grained counterparts with coarse-graining length $L = 5.9$ $\textrm \AA$, respectively.
  (e) Semi-logarithmic plot of the spatial autocorrelation function versus distance for $1/I_\textrm{BP}$ and $\Delta Q$. The red dashed lines are the best fits according to an empirical equation $C\left( r \right) = \exp \left( {{{ - r} / {\xi} }} \right)$. When $r$ = 5.9 $\textrm \AA$, $C\left( r \right)$ decays to approximately $\exp(-3)$, as shown by the blue dashed line.}
  \label{Fig:5}
\end{figure*}


\begin{figure}
  \centering
  \includegraphics[width=0.38\textwidth]{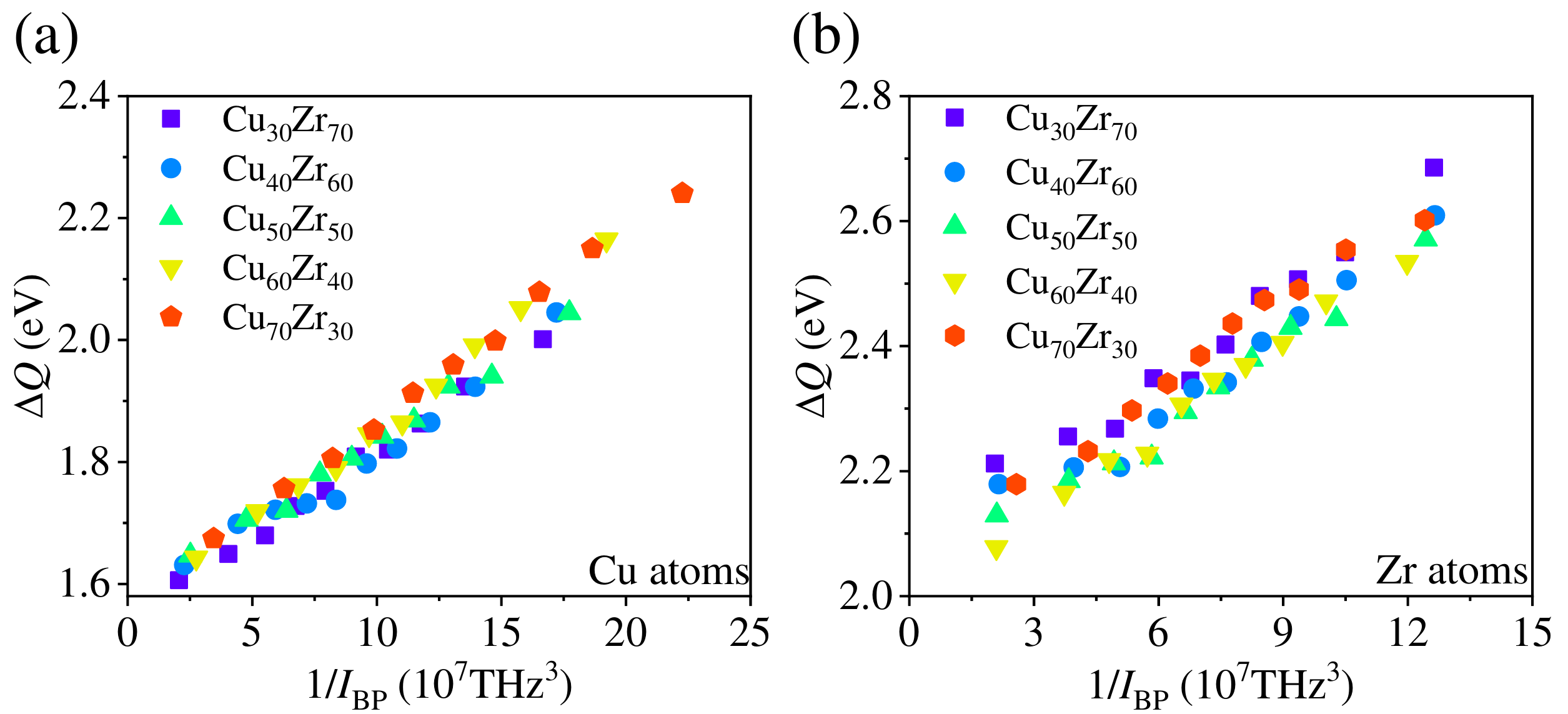}
  \caption{Robustness of the inverse proportionality between the activation energy and the intensity of boson peak against variation in chemical composition. The surveyed glass compositions include a set of Cu$_x$Zr$_{100-x}$ MGs ($x$ = 30, 40, 50, 60, 70). (a) for Cu atoms, and (b) for Zr atoms. }\label{Fig:6}
\end{figure}

\begin{figure}
  \centering
  \includegraphics[width=0.38\textwidth]{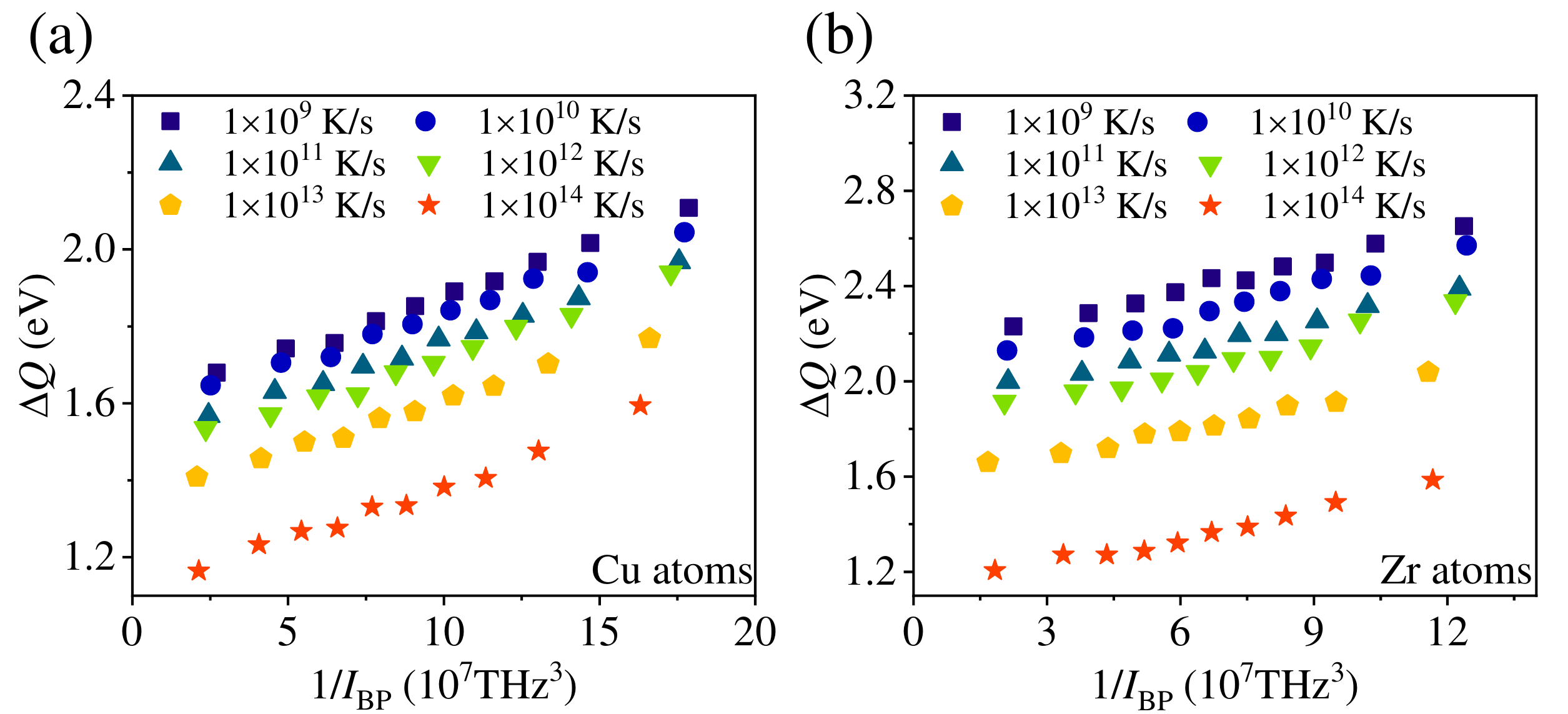}
  \caption{Robustness of the inverse proportionality between the activation energy and the intensity of boson peak against variation in cooling history. The cooling rates applied during sample preparation are varying from $1\times 10^9$ to $1\times 10^{14}$ K/s. (a) for Cu atoms, and (b) for Zr atoms.}\label{Fig:7}
\end{figure}

The $I_{\rm{BP}}$--$\Delta Q$ correlation can be studied in terms of spatial coarse-graining.  The spatial distributions of the particle-level intensity of boson peak $I_{\rm{BP}}$ and activation energy $\Delta Q$, as well as their coarse-grained counterparts, i.e., $[1/I_\textrm{BP}]_\textrm{CG}$ and $\Delta Q_\textrm{CG}$, are shown in Fig.\ref{Fig:5}(a)--(d). The 3D rendering is further provided in Figs. S2 and S3. All of them consistently imply that the high-$I_{\textrm{BP}}$ regions and the low-barrier regions overlap with each other significantly, once a certain (optimal) value of spatial coarse-graining length is chosen. Again, the best correlation is established with coarse-graining size of 5.9 $\textrm \AA$. This value arises as the characteristic decay length of the spatial autocorrelation function for both quantities. Here, the normalized spatial autocorrelation function of a physical entity is defined as
\begin{equation}\label{Eq:1}
  C\left( r \right) = \frac{{\left\langle {\Delta {P_{{r_0}}}\Delta {P_{{r_0} + r}}} \right\rangle  - {{\left\langle {\Delta {P_{{r_0}}}} \right\rangle }^2}}}{{\left\langle {\Delta {P_{{r_0}}}^2} \right\rangle  - {{\left\langle {\Delta {P_{{r_0}}}} \right\rangle }^2}}},
\end{equation}
where $P_{r_0}$ and $P_{r_0 + r}$ denote the values of the property ``$P$" at a reference position $r_0$, and that at a distance of $r$ from the reference atom, respectively. Here, $\Delta$ denotes the deviation from the ensemble averaged value. The operator $\left\langle  \cdots  \right\rangle$ represents the operation of ensemble average. As shown in Fig.\ref{Fig:5}(e), the spatial autocorrelation function for the particle-level intensity of the boson peak, and that for the activation energy, exhibit exactly the same decay with distance $r$, with exactly the same value of correlation length. When $r = 5.9$ $\textrm \AA$, $C\left(r\right)$ decays to approximately $\exp \left( -3 \right)$ of the reference value for both quantities, which corresponds to an optimal compromise, since lower values would imply losing the effect of medium-range correlations, whereas larger values would see the correlation vanish altogether.

As mentioned above, a spatial region with a high-boson-peak contribution can be used as predictor of a locally shallow basin with lower activation energy. The universality of this relationship is demonstrated by examining different glasses. As shown in Fig.\ref{Fig:6} and Fig.\ref{Fig:7}, such correlation is robust against variation in both composition and cooling history. Since low barriers are usually associated with asymmetry in PEL topology~\cite{Delaire2015}, one may deduce that there is intimate correlation between anharmonicity and boson peak in metallic glasses. On the contrary, large activation energy is linked with deeper valleys in PEL, which are usually well described by the harmonic transition state theory. In this sense, lower $\Delta Q$ means stronger deviation from harmonicity, hence ``soft'' and emergently anharmonic regions of the PEL. This idea will be justified in the following from different perspectives at the atomic scale.

\begin{figure*}
  \centering
  \includegraphics[width=0.75\textwidth]{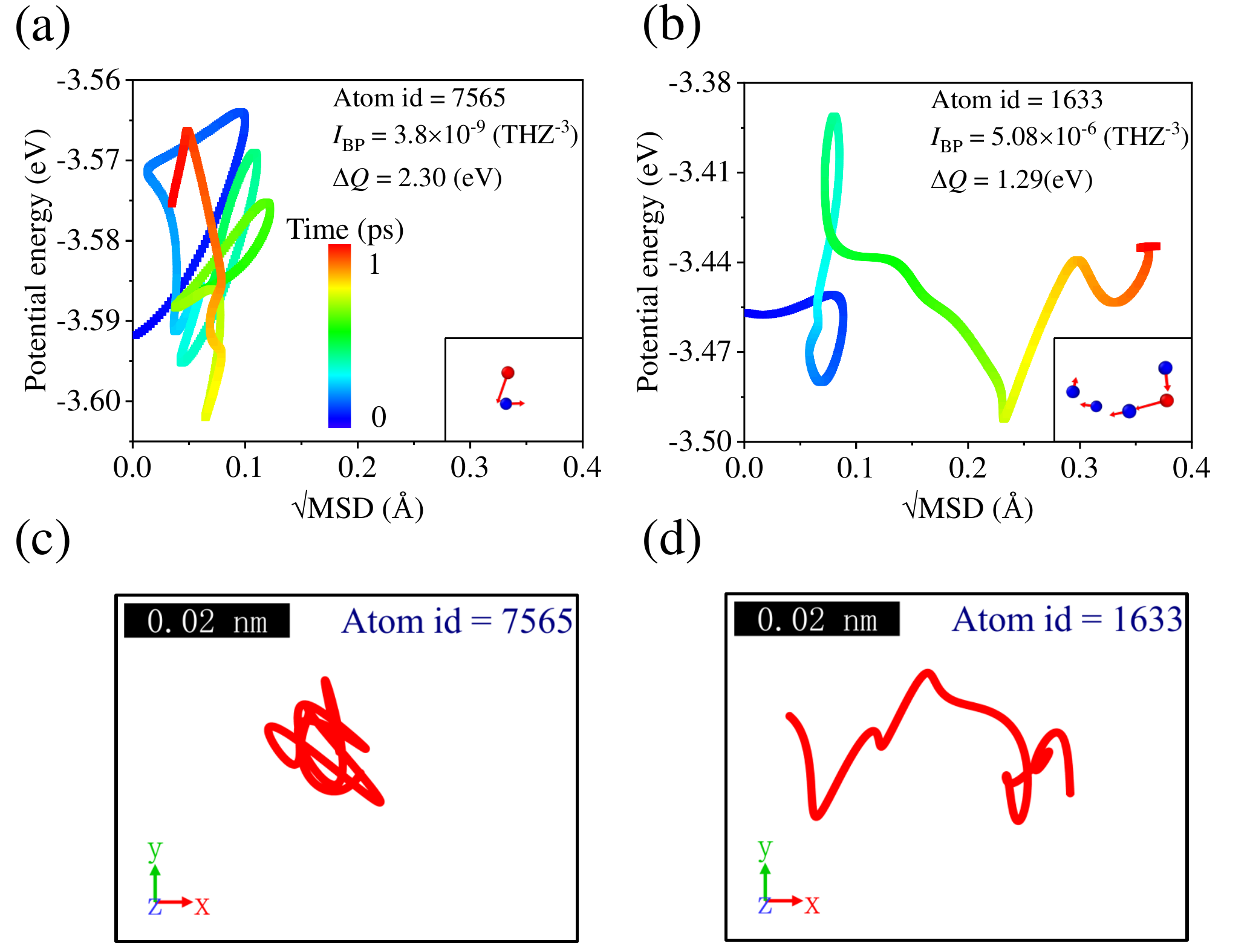}
  \caption{Demonstration of the anharmonicity of local energy basins.
  (a)--(b) Potential energy as a function of RMSD of the atoms with the weakest and strongest intensity of boson peak, respectively. The color bar corresponds to elapsing time. The insets show the pattern of atomic motions in terms of displacement vector. Only the atoms travelled longer than 1 $\textrm \AA$ are shown for clarity, with blue spheres representing copper atoms and red for zirconium ones. Higher boson-peak atoms usually experience string-like motion.
  (c)--(d) The trajectories of the atoms corresponding to (a) and (b).}
  \label{Fig:8}
\end{figure*}



\begin{figure}
  \centering
  \includegraphics[width=0.45\textwidth]{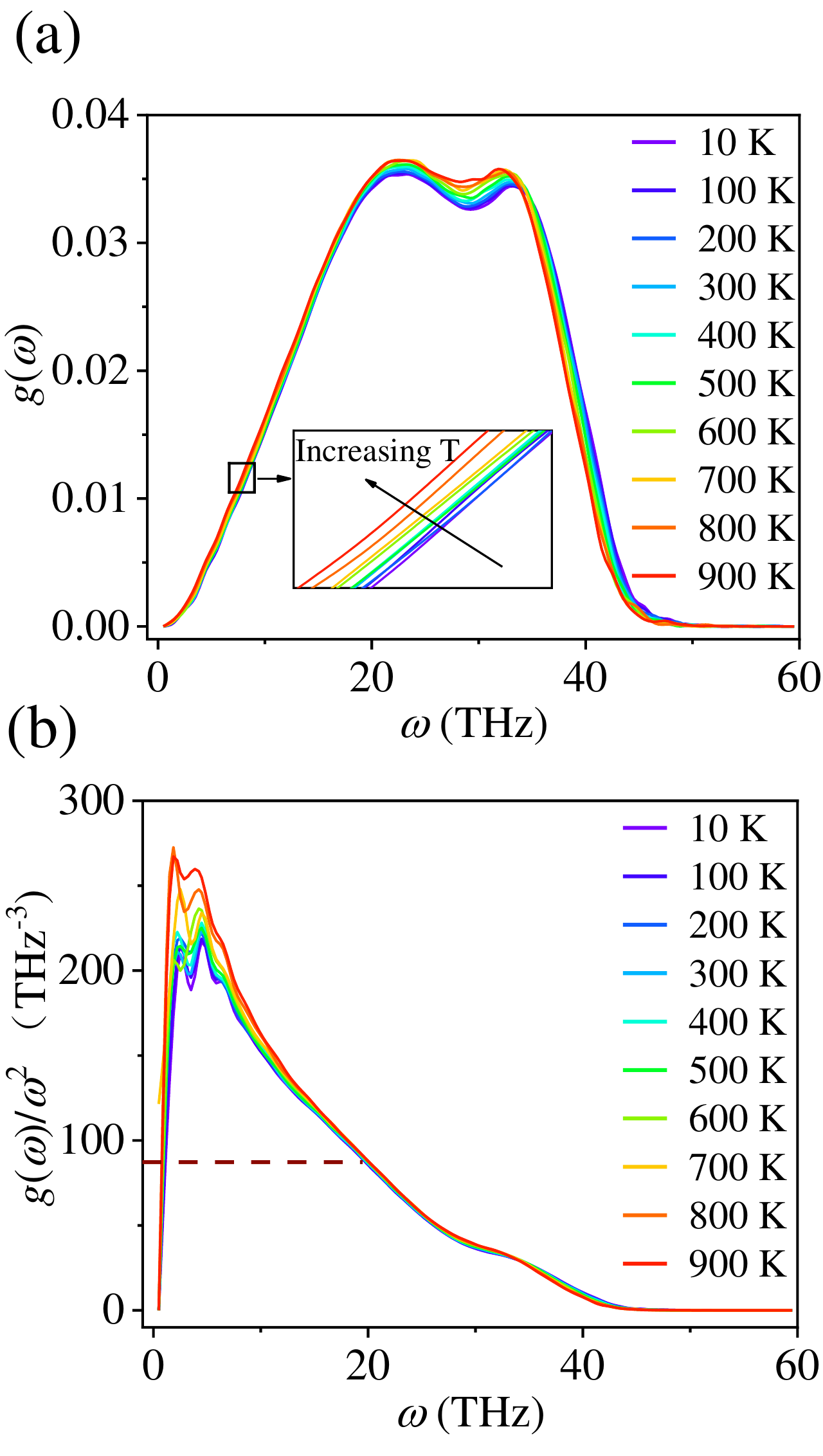}
  \caption{Temperature relevance of the boson peak.
  (a). VDOS of the glassy inherent structures at different temperatures. Inset indicates increasing vibrational soft modes at higher temperature.
  (b). Reduced VDOS by Debye-squared law. The dashed line shows the Debye level. Boson peak intensifies by increasing temperature.}
  \label{Fig:9}
\end{figure}


\begin{figure}
  \centering
  \includegraphics[width=0.45\textwidth]{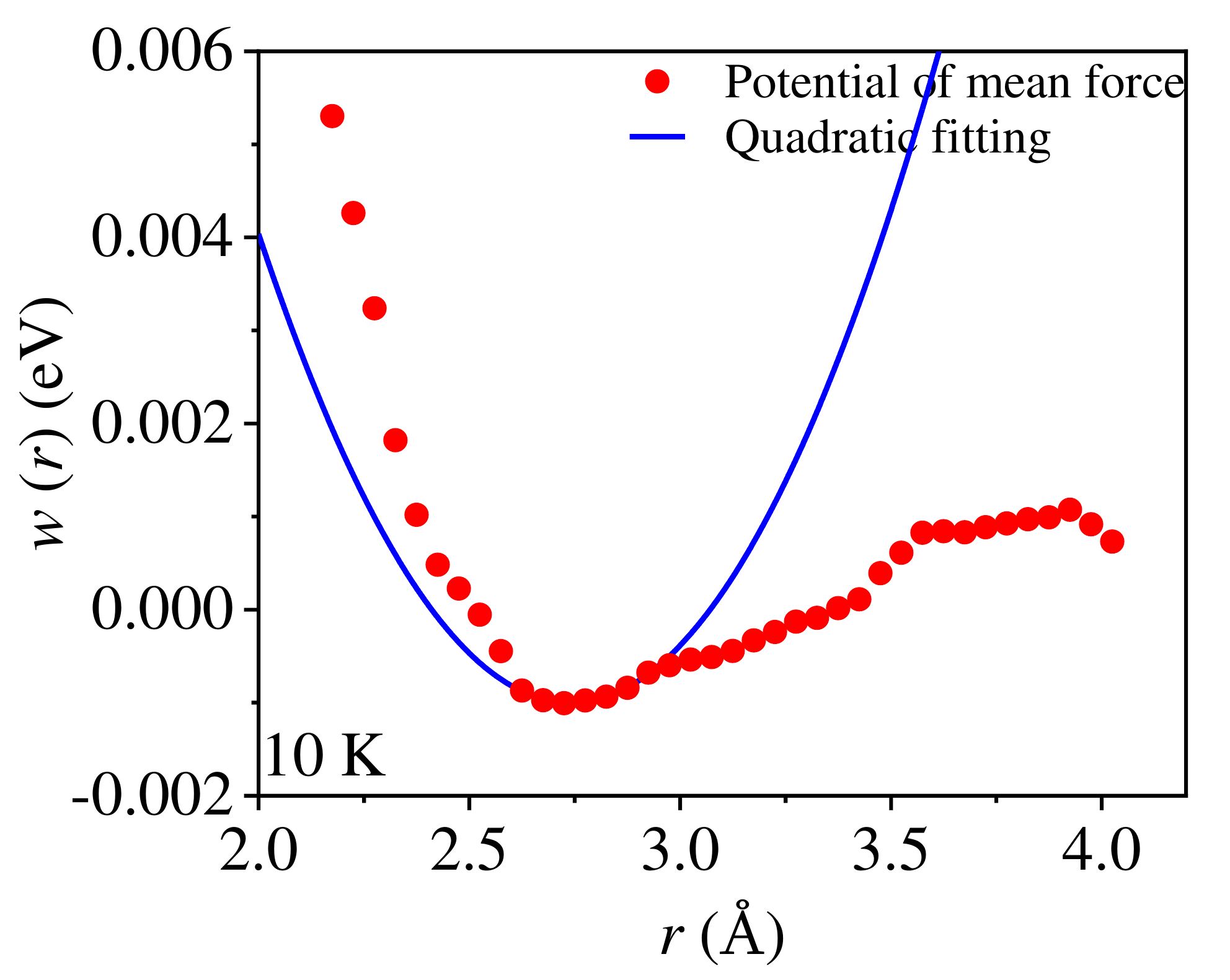}
  \caption{Potential of mean force as a function of distance. Scattered points in red are raw data. The blue line is the best quadratic fitting of points at the valley.}\label{Fig:10}
\end{figure}

\begin{figure*}
  \centering
  \includegraphics[width=0.75\textwidth]{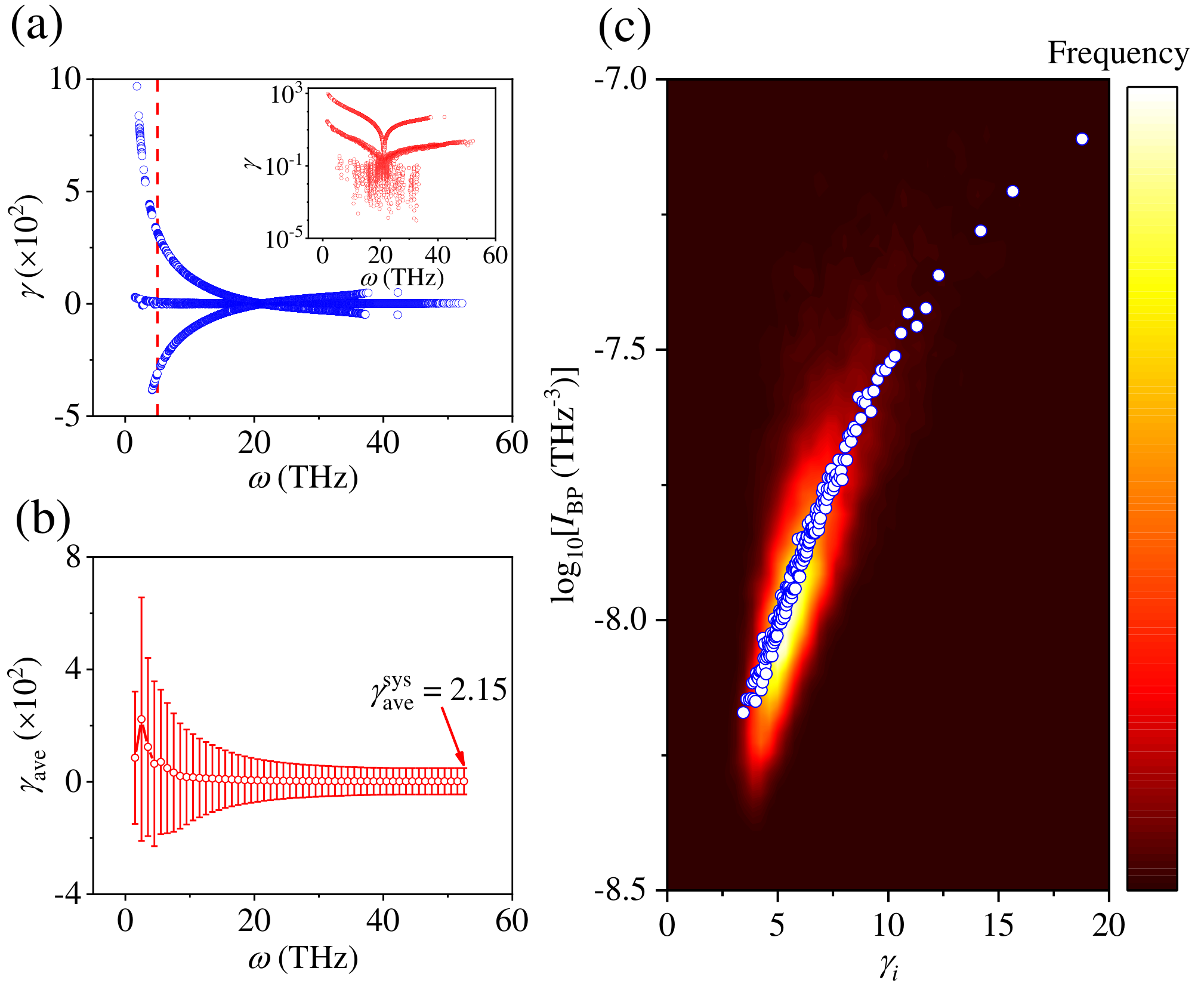}
  \caption{Anharmonicity unveiled by mode- and particle-level Gr\"{u}neisen parameter.
  (a). Mode Gr\"{u}neisen parameter as a function of phonon frequency, with the inset being the semi-logarithmic plot. The dashed vertical line indicates the position of boson peak frequency. Gr\"{u}neisen parameter is large at low frequencies, indicating a quasi-localized nature of the low-frequency vibration.
  (b). Mode averaged Gr\"{u}neisen parameter $\gamma_\textrm{ave}$ as a function of frequency. Error bars denote the standard deviation in the frequency domain below the shown value.
  (c). Semi-logarithmic plot of the intensity of single-particle boson peak as a function of the single-particle Gr\"{u}neisen parameter. The background heat map denotes the raw data colored by number density, while the scattered circles represent numerical coarse-grained values with binning size of a hundred atoms.}
  \label{Fig:11}
\end{figure*}


To demonstrate the anharmonicity of the energy basins which contribute to the BP, we run MD in Cu$_{50}$Zr$_{50}$ glass sample for 1 ps at 100 K, after sufficient thermal relaxation. The potential energy and the root mean-squared displacement (RMSD) of the system are recorded in Fig.\ref{Fig:8} as a function of time. We pick two extreme Cu atoms, with strongest and weakest  contribution to the boson peak, respectively, for demonstration. For the low-boson-peak atom with ID = 7565 shown in Fig.\ref{Fig:8}(a), each fragment of PEL closely resembles a quadratic function. However, the PEL fragment of the high-boson-peak atom with ID = 1633 shown in Fig.\ref{Fig:8}(b) strongly deviates from the harmonic approximation. This indicates that the atomic motions contributing to boson peak explore the more anharmonic topology of the local PEL. These atoms are found to move far away from their equilibrium positions, as further evidenced by the trajectory shown in Fig.\ref{Fig:8}(d). Instead, atoms that do not participate in the boson peak are limited in their motions to the harmonic basins of the PEL. Low-boson-peak atoms are found to vibrate near their local equilibrium position, as shown in Fig.\ref{Fig:8}(c). Moreover, it is interesting to find that the high-$I_\textrm{BP}$ experiences string-like collective motion, as shown in the inset of Fig.\ref{Fig:8}(b), in analogy to the atomic-scale pattern of $\beta$ relaxation \cite{Donati1998,PazminoBetancourt2015,Yu2017}. The local structural changes during such “chain-like” diffusive motion are further detailed in Fig. S4, which shows that a significant fraction of neighboring atoms around a high-boson-peak atom experience essential structural frustrations. It indicates that these atoms which contribute to the boson peak are not only elastically soft but plastically soft as well. This new link that we found between BP and stringlets aligns well with very recent results of Douglas and co-workers~\cite{Zhang2021}. Further, the generality of such emergently anharmonic dynamics of high-$I_\textrm{BP}$ atoms is verified by applying this analysis to more atoms in Figs. S5 and S6.

Temperature dependence provides further insights into the anharmonic contribution to the boson peak. If the local PEL were completely harmonic for a certain vibration mode, there would be no dependence on temperature. Here, we performed the vibrational analysis at different temperatures. To do this, VDOS was obtained by direct diagonalization of Hessian matrix of stable structures minimized from equilibrated configurations at different temperatures. Fig.\ref{Fig:9} shows the VDOS and its Debye-normalized at different temperatures. Increasing temperature leads to the shift of VDOS and boson peak towards smaller values of frequency. Meanwhile, the intensity of boson peak is enhanced as temperature increases in agreement with the behaviour seen for polymer glasses~\cite{Palyulin2018}. The phenomena are more conspicuous across the glass transition temperature, $T_\textrm{g} = 650$ K for Cu$_{50}$Zr$_{50}$ (The glass transition temperature is determined by examination of the discontinuity in the slope of volume-temperature curve upon cooling a glass-forming liquid to glass by MD). Thus, temperature-dependent VDOS points to intimate correlation between the emergent anharmonicity and the boson peak. It is of note that this is a mean-field-type scenario, which assumes the glass remains in the same energy basin without escaping upon rising temperature. If the inherent structure is changed during heating, then the physical picture becomes intricate due to existence of hierarchical PEL structures in the general glasses. Note that some early experiments observed that the Raman scattering intensity is proportional to $n \left( \omega \right) + 1$ \cite{Shuker,Leadbetter}. Therefore, the peak in the VDOS was concluded to be temperature independent. This is possibly caused by either 1) the Bose-Einstein exponential function at room temperature, which makes the temperature dependence of the BP not visible, or 2) experimental error bars, or a combination of 1) and 2).

Another evidence to intuitively demonstrate the emergent anharmonicity is the potential of mean force $w\left( r \right)$, which is defined as the reversible work to move two atoms from infinite distance to separation with distance $r$. This reversible work is the effective interaction in the mean-field sense since it denotes the change of free energy for the process and its gradient gives the average of the force over the equilibrium distribution of all other atoms. According to e.g. \cite{Chandler1986}, the exact definition of potential of mean force is $w\left( r \right) =  - {k_{\rm{B}}}T\ln g(r)$, where $g(r)$ denotes the radial distribution function. We test the idea in an inherent structure of 10 K with atoms interacting via the present empirical EAM potential. The result is shown in Fig.\ref{Fig:10}, in which the potential of mean force is plotted as a function of distance. The red scattered data represent EAM calculations and the blue line denotes the best quadratic fitting of points at the valley. It shows strong anharmonicity expressed by the strong deviation from the quadratic curve once the state leaves the potential bottom. Therefore, large values of cubic and quartic terms are expected in the Taylor expansion of the real interactive potential.

The final discussion, which further demonstrates and quantifies the emergent anharmonicity of boson peak vibrations, comes from the mode- and atomic-resolved Gr\"{u}neisen parameter $\gamma$, which is directly associated to the third- and higher-order anharmonic coefficients in the Taylor expansion of the interatomic interactions~\cite{Dolling1966,Krivtsov2011,Delaire2015,Setty2020}. The atomistic Gr\"{u}neisen parameter can straightforwardly quantify the level of anharmonicity of a local PEL fragment experienced by a given atom, furthermore the linewidth of acoustic phonons can be shown~\cite{Akhiezer,Tanaka} to be $\Gamma \sim Dq^{2}$, with the vibration diffusivity $D$ being proportional to the average Gr\"{u}neisen parameter $\gamma$ of the material at hand~\cite{Boemmel,Setty2020}. First of all, the mode-Gr\"{u}neisen parameter, $\gamma_k$, is calculated according to the derivative of the phonon frequency with respect to volume variation
\begin{equation}\label{Eq:3}
  {\gamma _k} =  - \frac{V}{{{\omega _k}}}\frac{{\partial {\omega _k}\left( V \right)}}{{\partial V}},
\end{equation}
where $V$ is the volume of the simulation box, and $\omega_k$ is the phonon frequency of the $k$th normal mode. It can be approximated by the numerical differentiation method \cite{Togo2015}:
\begin{equation}\label{Eq:4}
  {\gamma _k} =  - \frac{V}{{{\omega _k}}}\left\langle {\frac{{\Delta {\omega _k}\left( V \right)}}{{\Delta V}}} \right\rangle .
\end{equation}
To obtain the mode-Gr\"{u}neisen parameter, a three-step diagonalization of the Hessian matrix is carried out accounting for the variation of phonon frequency at 0.99 $V_0$, 1.0 $V_0$ and 1.01 $V_0$, respectively. $V_0$ is the volume of the system at ground state. As shown in Fig.\ref{Fig:11}(a), $\gamma_k$ is then plotted as a function of the phonon frequency. Surprisingly, huge values of the mode-Gr\"{u}neisen parameter appear at low frequencies. It should be noted that the apparent extreme values of the mode-Gr\"{u}neisen parameter corresponding to the low-frequency modes must originate from the local structural rearrangements (made possible due to low activation energies; see Fig.\ref{Fig:1}) upon volumetric expansion or contraction. Actually, even an extremely small volume change will cause some tiny variation of local inherent structures since there are certain abundant local soft structures similar to liquids with activation energies close to zero; see further evidence in Refs. \cite{Wei2019,Han2020}. However, it is such diffusive phonon that identifies the soft modes. It also generally indicates that the definition of the $k$th normal mode -- sorted by the magnitude of the normal mode frequency $\omega_k$ at ground state $V_0$ -- is not invariant against volumetric variation. This is in line with the work by Fabian and Allen \cite{Fabian1997,Fabian1999} which shows exactly similar trend of mode-Gr\"{u}neisen parameter vs mode frequency for amorphous silicon. Fig.\ref{Fig:11}(a) implies that the boson peak is correlated to the emergently anharmonic interactions, and that the energy flows from one mode to another, in analogy to the anharmonic/nonlinear Fermi-Pasta-Ulam problem. Moreover, the dashed vertical line in Fig.\ref{Fig:11}(a) indicates the frequency of boson peak, which appears right at the inflection point of the $\gamma$ vs $\omega$ function. The inset of Fig.\ref{Fig:11}(a) is a semi-logarithmic plot, which shows that a large part of data points for the mode-Gr\"{u}neisen parameter are near zero. Only a small fraction of modes contribute to the big absolute value of $\gamma_k$ and thus to the boson peak. The discontinuity in $\gamma_k$ in the inset can be attributed to the periodic boundary condition used in atomistic simulations.

Further, the anharmonic feature is demonstrated also at the system level. In Fig.\ref{Fig:11}(b),  $\gamma_{\textrm{ave}}$, which denotes the average value of the mode Gr\"{u}neisen parameter within $\omega < \omega_0$, is plotted as a function of frequency. Even on a system level, the average $\gamma_k$ has quite large values (in range between 1 and 2), suggesting emergently anharmonic effects, especially, at low frequencies. It is interesting to note that $\gamma_\textrm{ave} \left({\omega < \omega_0}\right)$ has a maximum at 2.5 THz, which is in accord with the position of boson peak (2.5--5.5 THz) as shown in Fig.\ref{Fig:9}(b).

The above picture, where anharmonicity dominates the THz and the lower-energy spectrum, whereas the high-energy spectrum is less anharmonic, is also in line with experiments by Monaco and co-workers on various glasses, which showed that anharmonic damping is active at low wavenumber, while the more ``harmonic'' Rayleigh  scattering/damping~\cite{Cui2019} dominates at higher frequencies~\cite{Baldi2014}.

Finally, to show the robustness of the correlation between intensity of boson peak and anharmonicity, we further define a particle-level Gr\"{u}neisen parameter $\gamma_i$ via summation of all contributions from individual mode-Gr\"{u}neisen parameters projected onto the polarization direction of a specific atom $i$. That is
\begin{equation}\label{Eq:5}
  {\gamma _i} = \sum\limits_k { - \frac{V}{{{\omega _k}}}\left\langle {\frac{{\Delta {\omega _k}\left( V \right)}}{{\Delta V}}} \right\rangle {{\left| {\textbf{e}_k^i} \right|}^2}} ,
\end{equation}
where $\textbf{e}_k$ is the eigenvector of the $k$th eigenmode, and $\textbf{e}_k^i$ denotes the corresponding polarization vector of the atom $i$ in the $k$th normal mode.
Fig.\ref{Fig:11}(c) shows the statistical $I_{\rm{BP}}$--$\gamma_i$ correlation. Here, the background shows the raw data colored by the number density, while the scattered points are the result of numerical coarse graining with proper bin size. Throughout the whole frequency domain, the particle-level intensity of boson peak reveals a strong positive correlation with the particle-level Gr\"{u}neisen parameter, with a robust exponential correlation. It should be noted that there is a considerable deviation from the exponential trend at high $\gamma_{i}$ values larger than 10. This is probably due to the change of inherent structure upon volumetric variation when the Gr\"{u}neisen parameter is estimated according to Eq. \ref{Eq:5}. Therefore, the correlation for $\gamma_{i} > 10$ is apparent and should not be understood as the actual value of Gr\"{u}neisen parameter for the softest modes. Generally, it is evident from this picture that the atoms with the largest atomic-level Gr\"{u}neisen parameter values contribute the most to the boson peak intensity.
Hence, also the particle-level information strongly indicates that the emergent anharmonicity plays an important role of the boson peak in disordered materials.

\section{Conclusions}
We presented a detailed quantitative characterization of atomic vibrations in atomic glasses from the point of view of emergent anharmonicity. In contradiction with the current dominant paradigm~\cite{Schirmacher2007,Schirmacher2013,Schirmacher2020} that postulates that the excess of vibrational modes in the THz range of glasses (known as boson peak anomaly) is due to ``harmonic'' dissipationless processes induced solely by disorder, we have demonstrated quantitatively at the atomistic level that the boson peak is emergently correlated with anharmonic vibrations. This is in line with early theoretical models~\cite{Buchenau1991,Buchenau1992,Gurevich1993} and confirms the universal  mechanism proposed in~\cite{Baggioli2019} for the origin of the boson peak due to the Ioffe-Regel crossover from ballistic to diffusive anharmonic propagation of vibrational excitations in glasses as well as in ordered crystals~\cite{Tamarit2017,Jezowski,Tamarit2019}. The above framework  provides a natural connection between PEL and vibrational eigenmodes in solid-state systems. The emerging picture of disorder and anharmonicity being the two sides of the same coin, opens up plenty of opportunities for structure-property relations and for material discovery in the area of amorphous materials for mechanical and thermal transport applications, as well as for crystalline strongly anharmonic thermoelectric materials~\cite{Delaire2011,Delaire2015,Delaire2020,Wolverton} where the boson peak plays an increasingly important role~\cite{Suekuni,Wolverton,Zaccone_thermoelectric,Marzari,Ren2021}.

\begin{acknowledgements}
This work was financially supported by National Natural science Foundation of China (Grant No. 12072344), and the Youth Innovation Promotion Association of Chinese Academy of Sciences (Grant No. 2017025). A.Z. acknowledges the financial support from US Army Research Office, Contract W911NF-19-2-0055.
\end{acknowledgements}

%

\end{document}